\documentclass[aps,prl,twocolumn,showpacs,superscriptaddress]{revtex4}
\usepackage{graphicx}% Include figure files
\usepackage{dcolumn}% Align table columns on decimal point
\usepackage{bm}% bold math
\usepackage{color}

\newcommand{\sqrtsNN}{\mbox{$\sqrt{s_{_{\mathrm{NN}}}}$}}
\newcommand{\dAu}{\textit{d}+Au}

\newcommand{\AuAu}{Au+Au}

\newcommand{\pp}{\mbox{\textit{p}+\textit{p}}}

\newcommand{\meanpt}{\mbox{$\langle p_T \rangle$}}
\newcommand{\pt}{\mbox{$p_T$}}

\newcommand{\gev}{\mbox{$\mathrm{GeV}$}}
\newcommand{\gevcc}{\mbox{$\mathrm{GeV/}c^2$}}

\newcommand{\gevc}{\mbox{${\mathrm{GeV/}}c$}}

\newcommand{\RAA}{\mbox{$R_{AuAu/dAu}$}}

\newcommand{\dedx}{\mbox{$dE/dx$}}

\newcommand{\nbin}{\mbox{$N_{\mathrm{bin}}$}}

\begin{document}

% A useful Journal macro
\def\Journal#1#2#3#4{{#1} {\bf #2}, #3 (#4)}

% Some useful journal names
\def\NCA{\em Nuovo Cimento}
\def\NIM{\em Nucl. Instr. Meth.}
\def\NIMA{{\em Nucl. Instr. Meth.} A}
\def\NPB{{\em Nucl. Phys.} B}
\def\NPA{{\em Nucl. Phys.} A}
\def\PLB{{\em Phys. Lett.} B}
\def\PRL{{\em Phys. Rev. Lett.}}
\def\PRC{{\em Phys. Rev.} C}
\def\PRD{{\em Phys. Rev.} D}
\def\ZPC{{\em Z. Phys.} C}
\def\JPG{{\em J. Phys.} G}
\def\EPJ{{\em Eur. Phys. J.} C}
\def\RPP{{\em Rep. Prog. Phys.}}
\def\IPA{{\em Int. J. Mod. Phys.} A}
\def\IPE{{\em Int. J. Mod. Phys.} E}
\def\JHEP{{\em J. High Energy Phys.}}

%\title{Scaling of charm integrated cross section and modification of its
%transverse momentum spectra in \dAu\ and \AuAu\ collisions at
%RHIC}
\title{Charmed hadron production at low transverse momentum in Au+Au collisions at RHIC}

\date{\today}
\affiliation{Argonne National Laboratory, Argonne, Illinois 60439}
\affiliation{University of Birmingham, Birmingham, United Kingdom}
\affiliation{Brookhaven National Laboratory, Upton, New York
11973} \affiliation{California Institute of Technology, Pasadena,
California 91125} \affiliation{University of California, Berkeley,
California 94720} \affiliation{University of California, Davis,
California 95616} \affiliation{University of California, Los
Angeles, California 90095} \affiliation{Carnegie Mellon
University, Pittsburgh, Pennsylvania 15213}
\affiliation{University of Illinois at Chicago, Chicago, Illinois
60607} \affiliation{Creighton University, Omaha, Nebraska 68178}
\affiliation{Nuclear Physics Institute AS CR, 250 68
\v{R}e\v{z}/Prague, Czech Republic} \affiliation{Laboratory for
High Energy (JINR), Dubna, Russia} \affiliation{Particle Physics
Laboratory (JINR), Dubna, Russia} \affiliation{University of
Frankfurt, Frankfurt, Germany} \affiliation{Institute of Physics,
Bhubaneswar 751005, India} \affiliation{Indian Institute of
Technology, Mumbai, India} \affiliation{Indiana University,
Bloomington, Indiana 47408} \affiliation{Institut de Recherches
Subatomiques, Strasbourg, France} \affiliation{University of
Jammu, Jammu 180001, India} \affiliation{Kent State University,
Kent, Ohio 44242} \affiliation{Institute of Modern Physics,
Lanzhou, China} \affiliation{Lawrence Berkeley National
Laboratory, Berkeley, California 94720} \affiliation{Massachusetts
Institute of Technology, Cambridge, MA 02139-4307}
\affiliation{Max-Planck-Institut f\"ur Physik, Munich, Germany}
\affiliation{Michigan State University, East Lansing, Michigan
48824} \affiliation{Moscow Engineering Physics Institute, Moscow
Russia} \affiliation{City College of New York, New York City, New
York 10031} \affiliation{NIKHEF and Utrecht University, Amsterdam,
The Netherlands} \affiliation{Ohio State University, Columbus,
Ohio 43210} \affiliation{Panjab University, Chandigarh 160014,
India} \affiliation{Pennsylvania State University, University
Park, Pennsylvania 16802} \affiliation{Institute of High Energy
Physics, Protvino, Russia} \affiliation{Purdue University, West
Lafayette, Indiana 47907} \affiliation{Pusan National University,
Pusan, Republic of Korea} \affiliation{University of Rajasthan,
Jaipur 302004, India} \affiliation{Rice University, Houston, Texas
77251} \affiliation{Universidade de Sao Paulo, Sao Paulo, Brazil}
\affiliation{University of Science \& Technology of China, Hefei
230026, China} \affiliation{Shanghai Institute of Applied Physics,
Shanghai 201800, China} \affiliation{SUBATECH, Nantes, France}
\affiliation{Texas A\&M University, College Station, Texas 77843}
\affiliation{University of Texas, Austin, Texas 78712}
\affiliation{Tsinghua University, Beijing 100084, China}
\affiliation{Valparaiso University, Valparaiso, Indiana 46383}
\affiliation{Variable Energy Cyclotron Centre, Kolkata 700064,
India} \affiliation{Warsaw University of Technology, Warsaw,
Poland} \affiliation{University of Washington, Seattle, Washington
98195} \affiliation{Wayne State University, Detroit, Michigan
48201} \affiliation{Institute of Particle Physics, CCNU (HZNU),
Wuhan 430079, China} \affiliation{Yale University, New Haven,
Connecticut 06520} \affiliation{University of Zagreb, Zagreb,
HR-10002, Croatia}

\author{B.I.~Abelev}\affiliation{University of Illinois at Chicago, Chicago, Illinois 60607}
\author{M.M.~Aggarwal}\affiliation{Panjab University, Chandigarh 160014, India}
\author{Z.~Ahammed}\affiliation{Variable Energy Cyclotron Centre, Kolkata 700064, India}
\author{B.D.~Anderson}\affiliation{Kent State University, Kent, Ohio 44242}
\author{D.~Arkhipkin}\affiliation{Particle Physics Laboratory (JINR), Dubna, Russia}
\author{G.S.~Averichev}\affiliation{Laboratory for High Energy (JINR), Dubna, Russia}
\author{Y.~Bai}\affiliation{NIKHEF and Utrecht University, Amsterdam, The Netherlands}
\author{J.~Balewski}\affiliation{Indiana University, Bloomington, Indiana 47408}
\author{O.~Barannikova}\affiliation{University of Illinois at Chicago, Chicago, Illinois 60607}
\author{L.S.~Barnby}\affiliation{University of Birmingham, Birmingham, United Kingdom}
% ...

\collaboration{STAR Collaboration}\noaffiliation
\begin{abstract}

We report measurements of charmed hadron production from hadronic
($D\rightarrow K\pi$) and semileptonic ($\mu$ and $e$) decays in
200 \gev\ \AuAu\ collisions at RHIC. Analysis of the spectra
indicates that charmed hadrons have a different radial flow
pattern from light or multi-strange hadrons. Charm cross sections
at mid-rapidity are extracted by combining the three independent
measurements, covering the transverse momentum range that
contributes to $\sim$90\% of the integrated cross section. The
cross sections scale with number of binary collisions of the
initial nucleons, a signature of charm production exclusively at
the initial impact of colliding heavy ions. The implications for
charm quark interaction and thermalization in the strongly
interacting matter are discussed.

\end{abstract}

\maketitle

%%%%%%%%%%%%%%%%%%%%%%%%%%%%%%%%%%%%%%%%%%%%%%%%%%%%%%%%%%%%%%%%%%%%%%
% INTRODUCTION
%%%%%%%%%%%%%%%%%%%%%%%%%%%%%%%%%%%%%%%%%%%%%%%%%%%%%%%%%%%%%%%%%%%%%%

Charm quarks are a unique tool to probe the partonic matter
created in relativistic heavy-ion collisions at RHIC energies. Due
to their large mass ($\simeq$1.3 \gevcc), charm quarks are
predicted to lose less energy than light quarks by gluon radiation
in the medium~\cite{dead}. In contrast, recent measurements of the
\pt\ distributions and nuclear modification factors of
non-photonic electrons (NPE) from heavy quark decays at high \pt\
show a suppression level similar to light
hadrons~\cite{starphenixcharmraa}. This observation renews the
interest in charm production and the interactions of heavy charm
quarks with the hot and dense matter produced in nuclear
collisions at RHIC.

Measurements of charm production at low \pt, in particular radial
and elliptic flow, probe the QCD medium at thermal scales and are
thus sensitive to bulk medium properties like density and the drag
constant or viscosity. Model treatments for low-\pt\ charm
production, such as energy loss by collisional
dissociation~\cite{Ivancoll} and in-medium transport using a
diffusion formalism (in analog to Brownian motion) and resonance
cross sections~\cite{RappRaa}, can be used to infer transport
properties such as interaction cross sections and the medium
density. Calculations of charm transport in strongly coupling
theories using AdS/CFT correspondence~\cite{adscft} may allow to
further determine the viscosity or drag coefficient of quark-gluon
matter formed at RHIC. Ultimately, charm quark radial flow may
help establish whether light quarks thermalize~\cite{teaney}.

%Model calculations in an AdS/CFT correspondence or incorporating
%in-medium charm resonances/diffusion or collisional dissociation
%in a strongly interacting
%medium~\cite{teaney,RappRaa,Ivancoll,adscft} can be reasonably
%applied to describe charm spectra down to low \pt, and to obtain
%medium properties such as drag constant and viscosity. Heavy charm
%quarks can acquire flow from interactions with the dense medium of
%light quarks in analog to Brownian motion~\cite{teaney}. Charm
%decoupling from the medium also depends on the interaction
%cross sections and the medium density. Thus measurements of charm
%freeze-out properties, in particular radial and elliptic flow are
%vital to test light flavor thermalization and the early stage
%partonic density~\cite{teaney}.

%These postulations and predictions
These considerations often assume that charm quarks are produced
only in the early stages and their production rate is reliably
calculable by perturbative QCD~\cite{lin,cacciari}. Studies of the
binary collision (\nbin\ from Glauber model) scaling of the total
charm cross section from \dAu\ to \AuAu\ collisions can be used to
test these assumptions and determine if charm is indeed a good
probe with well-defined initial states. The total charm production
cross section is also an important input in models of $J/\psi$
production via charm quark coalescence in a Quark Gluon
Plasma~\cite{pbm}.

In this paper, we present the study of charm via measurements of
very low \pt\ muons~\cite{ffcharm}($0.17\leq\pt\leq0.25$ \gevc),
$D^{0}\rightarrow K\pi$ at low \pt\ ($\pt\leq2$ \gevc), and NPE
($0.9\leq\pt\leq5$ \gevc). The combination of these three
techniques provides the most complete kinematic coverage
($\sim$90\%) to date for charm production measurements at RHIC.
They also allow the study of the charmed hadron spectral shape in
order to explore the possibility of charm radial flow in \AuAu\
collisions.

%%%%%%%%%%%%%%%%%%%%%%%%%%%%%%%%%%%%%%%%%%%%%%%%%%%%%%%%%%%%%%%%%%%%%%%
%EXPERIMENT
%%%%%%%%%%%%%%%%%%%%%%%%%%%%%%%%%%%%%%%%%%%%%%%%%%%%%%%%%%%%%%%%%%%%%%

The data used for these analyses were taken with the STAR
experiment~\cite{STARWP} during the $\sqrtsNN=200$ \gev\ \AuAu\
run in 2004 at RHIC. The most central 0$-$80\% of the total
hadronic cross section as selected using the uncorrected charged
particle multiplicity at mid-rapidity ($|\eta|<0.5$), is used for
the minimum bias (minbias) measurement. A total of 13.3 million
minbias triggered events were used for the $D^0$ reconstruction. A
separate sample of central events was taken using an online
selection of the 0$-$12\% most central events based on the energy
deposited in the two Zero-Degree Calorimeters~\cite{STAR}. For
muons and NPE analysis, 15 million central (0$-$12\%) and 7.8
million minbias triggered events are used. All measurements are
presented as an average of particle and anti-particle yields at
mid-rapidity ($|y|<1$ for $D^{0}$ and $-1<y<0$ for $\mu$($e$)
limited by the TOF detector acceptance.).

\begin{figure}[tb]
\includegraphics[width=0.5\textwidth]{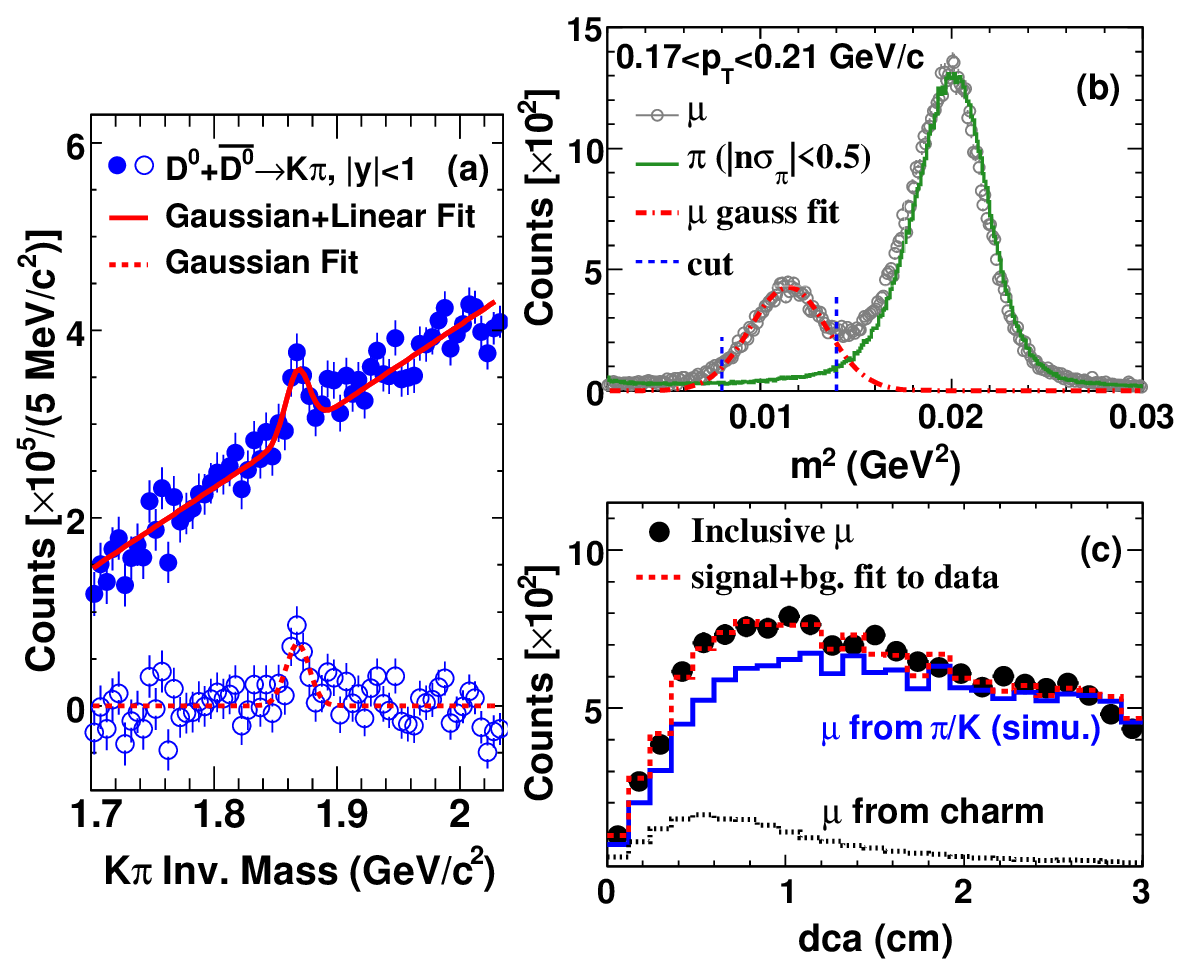}\vspace{-0.25cm}
\caption{(a) Kaon-pion pair invariant mass distribution for
minbias \AuAu\ collisions, after subtraction of mixed event
background (solid circles) and additional subtraction of a linear
residual background (open circles). A $4\sigma$ signal is
observed. (b) Mass squared distribution
$m^{2}=(p/\beta/\gamma)^{2}$ from TOF. (c) $\mu$ DCA distributions
(open circles). Histograms indicate the background from $\pi$/K
decays (solid curve), the contribution from prompt muons (dotted
curve) and the sum (dashed curve).}
\vspace{-0.35cm}\label{fig:figure1}
\end{figure}

%%%%%%%%%%%%%%%%%%%%%%%%%%%%%%%%%%%%%%%%%%%%%%%%%%%%%%%%%%%%%%%%%%%%%%
% ANALYSIS
%%%%%%%%%%%%%%%%%%%%%%%%%%%%%%%%%%%%%%%%%%%%%%%%%%%%%%%%%%%%%%%%%%%%%%
$D^0$ mesons were reconstructed through their decay
$D^0(\bar{D^0})\rightarrow K^{\mp}\pi^{\pm}$ with a branching
ratio of 3.83\%. The analysis is identical to that used for \dAu\
collisions~\cite{dAuCharm}. An example invariant mass distribution
after subtraction of the combinatorial background from mixed
events is shown in Fig.~\ref{fig:figure1}(a) (full circles). The
$D^0$ yield is extracted by fitting a Gaussian peak plus a linear
or second-order polynomial to describe the residual background to
the measured distribution (red curve in Fig.~\ref{fig:figure1}
(a)). The total systematic uncertainty on $D^0$ yield bin-by-bin
is $\sim$40$-$50\%, evaluated by varying the particle
identification conditions and yield extraction procedures.

Muons were identified by combining the Time Of Flight (TOF) and
ionization energy loss (\dedx) measured in the Time Projection
Chamber (TPC)~\cite{STAR,dAuCharm,ffcharm}.
Fig.~\ref{fig:figure1}(b) shows the $m^2$ distribution from TOF
for tracks with $0.17\leq\pt\leq0.21$ \gevc. Muons are selected in
the range $0.008\leq m^{2}\leq0.014$ GeV$^{2}/c^{4}$. The
distribution of the distance of closest approach (DCA) of the
tracks to the primary vertex is used to further separate $\mu$
from charm decay and from pion and kaon decays.
Fig.~\ref{fig:figure1}(c) shows the DCA distribution of $\mu$
after a statistical subtraction of the DCA distribution from
misidentified pions~\cite{muana}. The remaining dominant
background $\mu$ from $\pi/K$ weak decays have different DCA
distributions from the prompt muons, shown in
Fig.~\ref{fig:figure1}(c). The prompt $\mu$ raw yield was obtained
from a fit to the $\mu$ DCA distributions with the background DCA
distribution and the primary particle DCA
distribution~\cite{ffcharm}. Other sources of background
($\rho\rightarrow\mu^+\mu^-$, $\eta\rightarrow\gamma\mu^+\mu^-$,
$K^{0}_{S}\rightarrow\pi\mu\nu$, etc.) are found to be negligible
from simulations using yields from measured spectra at RHIC. The
systematic uncertainty is dominated by the uncertainty on the pion
contamination and was estimated using different \dedx\ cuts for
the pions.

%--===============================================================
%--===============================================================

%--===============================================================
\begin{figure}[tb]
\includegraphics[width=0.42\textwidth]{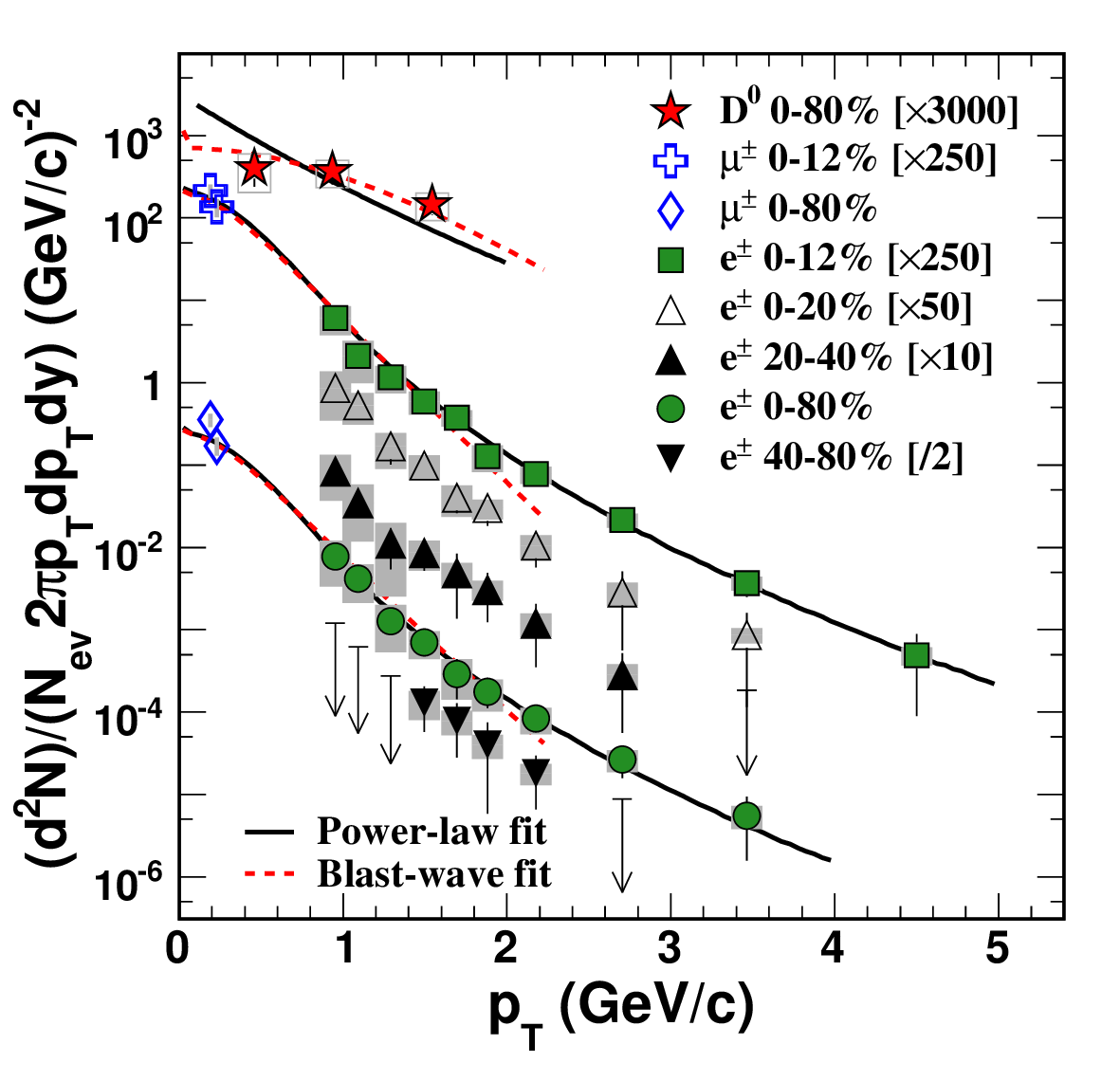}\vspace{-0.25cm}
\caption{\pt\ distributions of invariant yields for $D^0$,
charm-decayed prompt $\mu$ and NPE in different centralities.
Solid curves are power-law combined fit for $D^0$ and leptons.
Dashed curves are blast-wave fit. The gray bands are bin-to-bin
systematic uncertainties.} \vspace{-0.35cm}\label{fig:figure2}
\end{figure}
%--===============================================================

Electrons are identified up to $\pt=5$ \gevc\ by \dedx\ from TPC
after applying a TOF selection
$|1/\beta-1|\leq0.03$~\cite{STAR,dAuCharm}. The 5$-$15\%
bin-by-bin systematic uncertainties in inclusive electron yields
are dominated by the uncertainties in the raw yield extraction
using different fit functions. The dominant sources of photonic
electron background are photon conversions and $\pi^0(\eta)$
Dalitz decays. $e^+e^-$ pairs from these background sources were
subtracted using an invariant mass
technique~\cite{johnson,dAuCharm,ffcharm}. The photonic background
reconstruction efficiency was calculated in a Monte Carlo
simulation of $\pi^{0}$ and direct photons according to the
measured $\pi^+$~\cite{lqeloss} and $\pi^0$~\cite{pi0PHENIX}
spectra, and varies from 25\% at low \pt\ to 60\% at highest \pt\
in the range studied. The systematic uncertainties (7$-$16\%) for
photonic electrons are mainly from the combinatorial background
uncertainties and the background reconstruction efficiency. The
remaining background from photonic decays of $\eta, \omega, \rho,
\phi$ and $K$ was determined to be ${}^{<}_{\sim}5\%$ from
simulations and is subtracted in the final result. The ratio of
inclusive electrons to photonic background increases from $\sim1$
to 1.45 within the \pt\ range studied.

%%%%%%%%%%%%%%%%%%%%%%%%%%%%%%%%%%%%%%%%%%%%%%%%%%%%%%%%%%%%%%%%%%%%%%
% RESULT
%%%%%%%%%%%%%%%%%%%%%%%%%%%%%%%%%%%%%%%%%%%%%%%%%%%%%%%%%%%%%%%%%%%%%%

Fig.~\ref{fig:figure2} shows the \pt\ spectra for $D^0$, $\mu$ and
NPE for \AuAu\ minbias events. Additional centralities are shown
for the semi-leptonic decay measurements. We start with a \pt\
spectrum function for $D^0$ (power-law or blast-wave functions)
and obtain spectrum function from semileptonic decay for NPE and
$\mu$ from that. Additional \pt-dependent factor (upper and lower
bounds from FONLL~\cite{cacciari}) is applied to the lepton
function to take into account the contribution from bottom decays.
These functions are used in a combined fit~\cite{fitBR} to obtain
the mid-rapidity charm yield and $\langle p_T\rangle$. The
$\chi^2$ from the fit was with systematic errors included. The
correlations between systematic errors on different data points
were also taken into account in the combined fit~\cite{PDGerr}.
The $D^0$ \meanpt\ as calculated from power law fits are
0.92$\pm$0.06$(stat.)$$\pm$0.12$(sys.)$ \gevc\ for minbias and
0.95$\pm$0.04$\pm$0.16 \gevc\ for 0$-$12\% central \AuAu\
collisions.

\begin{figure}[tb]
\includegraphics[width=0.42\textwidth]{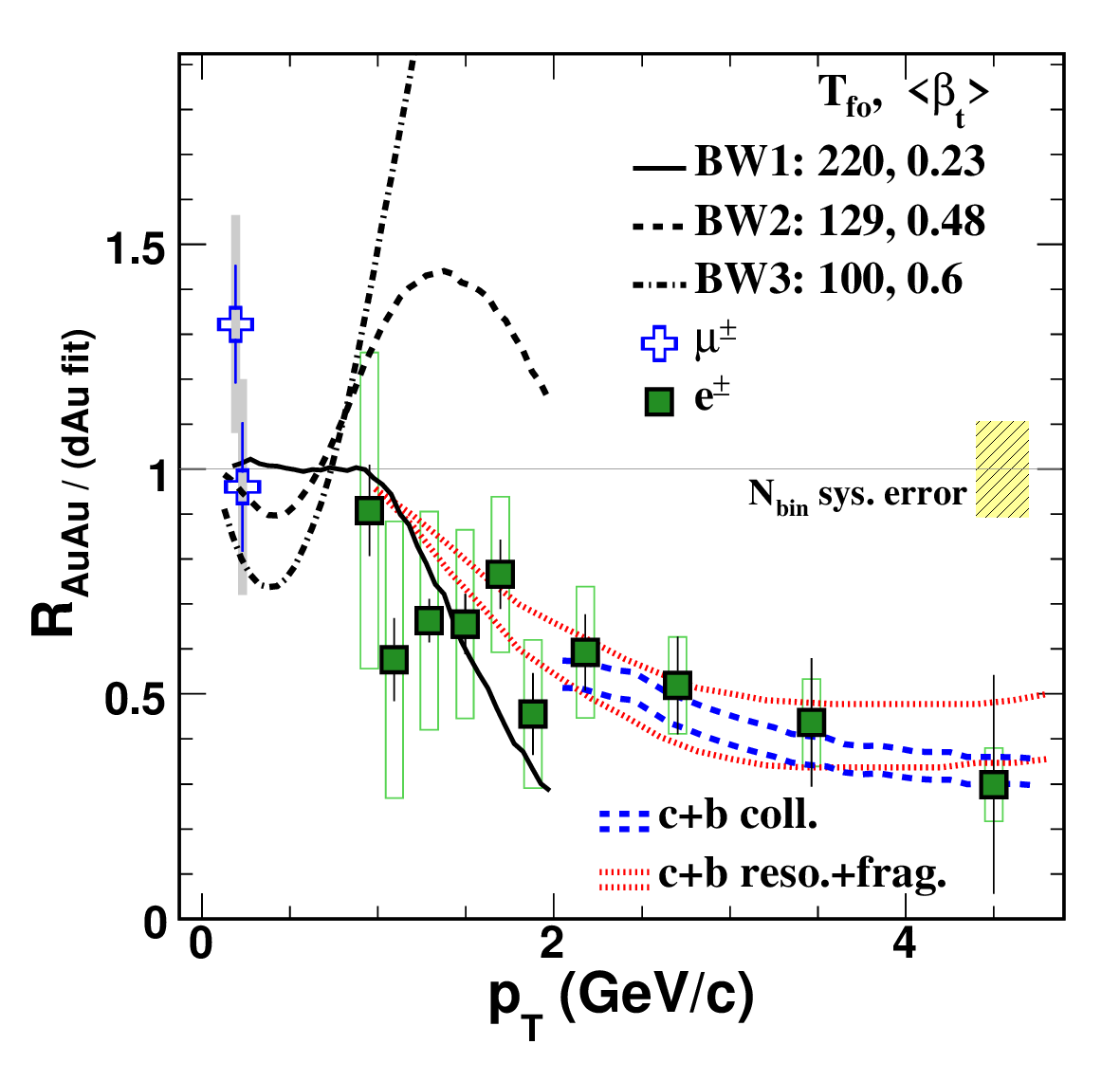}\vspace{-0.25cm}
\caption{Nuclear modification factor (\RAA) for 0$-$12\% \AuAu\
collisions. Bin-to-bin systematic uncertainties are represented by
the gray bands for muons and the open boxes for electrons. The
shaded band at unity shows the common $N_{bin}$ normalization
uncertainties. Model calculations are presented: coalescence and
fragmentation~\cite{RappRaa} (double-dotted curves), and
collisional dissociation of heavy meson~\cite{Ivancoll}
(double-dashed curves). The curves at $\pt\leq2$ \gevc\ indicate
blast-wave model with different freeze-out parameters ($T_{fo}$ in
MeV, $\langle \beta_t \rangle$) as indicated in the legend.}
\vspace{-0.35cm}\label{fig:figure3}
\end{figure}

The nuclear modification factors (\RAA)~\cite{star130auau} for
$\mu$ and NPE are shown in Fig.~\ref{fig:figure3}. The \RAA\ are
the ratios of the \pt\ spectra in \AuAu\ and in \dAu\ collisions
appropriately scaled with the number of binary collisions. No muon
measurements were carried out for the \dAu\ system. The \dAu\
reference is the decay lepton curve from the power-law combined
fit ($\langle p_T\rangle=1.18\pm0.02\pm0.10~\gevc,
n=11.5\pm0.5\pm1.5$) with $D^0$ and NPE from previously published
data~\cite{starphenixcharmraa,dAuCharm}. The nuclear modification
factor at low \pt\ is consistent with 1 to within the 25\%
relative uncertainty of the muon measurement, and then reduces at
higher \pt, reaching a value similar to light hadrons at high
\pt${}^{>}_{\sim}4$ \gevc~\cite{lqeloss}. Our result is consistent
with other measurements of NPE at high
\pt~\cite{starphenixcharmraa}. Model calculations incorporating
in-medium charm resonances/diffusion or collisional dissociation
in a strongly interacting medium~\cite{teaney,RappRaa,Ivancoll},
which can reasonably be applied to describe the NPE spectra down
to low \pt, are shown in Fig.~\ref{fig:figure3}.
%%%%%%%%%%%%%%%%%%%%%%%%%%%%%%%%%%%%%%%%%%%%%%%%%%%%%%%%%%%%%%%%%%%%%%
% DISCUSSION/THEORY
%%%%%%%%%%%%%%%%%%%%%%%%%%%%%%%%%%%%%%%%%%%%%%%%%%%%%%%%%%%%%%%%%%%%%%

To study whether charmed hadrons have similar radial flow to light
hadrons, we have included curves for the expected nuclear
modification factor from a blast-wave model, using the freeze-out
parameters for light hadrons~\cite{thermalhadron} (BW3 in
Fig.~\ref{fig:figure3}) and multi-strange hadrons~\cite{hyperon}
(BW2). The data and best blast-wave fit (BW1) show large
deviations from both these curves for $p_{T}>1$ \gevc, which
suggests that the charmed hadron freeze-out and flow are different
from light hadrons. We scanned the parameters to a 2-dimensional
$T_{fo}$, $\langle \beta_t \rangle$ space, the results show little
sensitivity to freeze-out temperature, but disfavor large radial
flow. These findings, together with the observation of large charm
elliptic flow~\cite{phenixcharmxsec}, are consistent with the
recent prediction from hydrodynamics~\cite{hirano}: elliptic flow
is built up at partonic stage, and radial flow dominantly comes
from hadronic scattering at later stage where charm may have
already decoupled from the system.

%These measurements can be used to study whether charmed hadrons
%have similar radial flow to light hadrons. The charm \meanpt\
%decreases from d+Au to Au+Au collisions while that of other
%hadrons including $\Omega$ increases sequentially from p+p to d+Au
%to Au+Au. This strongly suggests that the charmed hadron
%freeze-out and flow are different from light hadrons. To
%illustrate this further, we have included curves for the expected
%nuclear modification factor from a blast-wave model, using the
%freeze-out parameters for light hadrons~\cite{thermalhadron} (BW4
%in Fig.~\ref{fig:figure3}) and multi-strange
%hadrons~\cite{hyperon} (BW3). The charm measurement shows large
%deviations from both these curves. Our measurements show little
%sensitivity to freeze-out temperature, but disfavor large radial
%flow. These findings, together with the observation of large charm
%elliptic flow~\cite{phenixcharmxsec}, are consistent with the
%recent prediction from hydrodynamics~\cite{hirano}: elliptic flow
%is built up at partonic stage, and radial flow dominantly comes
%from hadronic scattering at later stage where charm may have
%already decoupled from the system.
%Blast-wave parameters with low temperature and moderate
%radial flow(dotted curve), or with high temperature and low radial
%flow (dashed curve) can describe our results.
%--====================================================================
%--====================================================================

\begin{figure}[tb]
\includegraphics[width=0.42\textwidth]{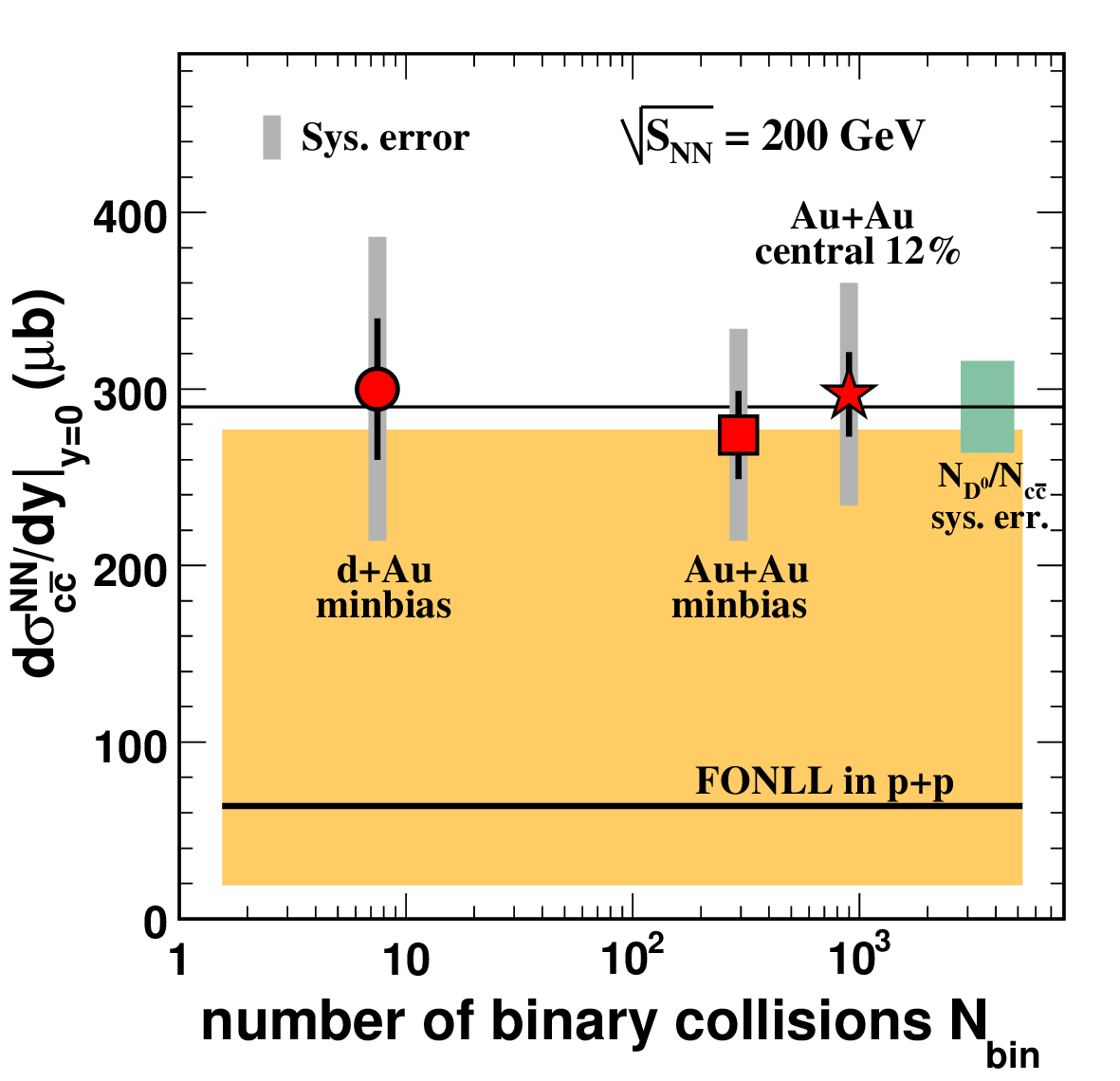}\vspace{-0.25cm}
\caption{Mid-rapidity charm cross section per nucleon-nucleon
collision as a function of $N_{bin}$ in \dAu, minbias and 0$-$12\%
central \AuAu\ collisions. The solid line indicates the average.
FONLL prediction is shown as a band around the central value
(thick line)~\cite{vogt}. }\vspace{-0.35cm}\label{fig:figure4}
\end{figure}

A combined fit to the NPE and $\mu$ spectra was used to obtain the
mid-rapidity charm cross section per nucleon-nucleon collision
($d\sigma_{c\bar{c}}^{NN}/dy$), shown in Fig.~\ref{fig:figure4}.
Yields are an average from the blast-wave and the power-law
functions, which are consistent within $\pm10\%$. For minbias
\AuAu\ collisions, the $D^0$ data were included in the fit,
assuming $N_{D^0}/N_{c\bar{c}}=0.54$$\pm$0.05~\cite{PDGerr} as is
seen in \pp\ collisions. Fig.~\ref{fig:figure4} shows
$d\sigma_{c\bar{c}}^{NN}/dy$ as a function of \nbin\ for minbias
\dAu, minbias and central \AuAu\ collisions. The charm production
cross section at mid-rapidity scales with number of binary
interactions from \dAu~\cite{dAuCharm} to central \AuAu\
collisions. The quality of this scaling can be quantified by the
slope of $d\sigma_{cc}/dy$ vs \nbin\
($d\overline{\sigma_{cc}^{NN}}/dy$), which is 290 $\mu b$ with
$\pm$15\% uncorrelated uncertainty. This indicates that charm
quarks are produced in the early stage of relativistic heavy-ion
collisions. The total cross section $\sigma_{c\bar{c}}^{NN}$
(extrapolated to the full rapidity using the rapidity distribution
from PYTHIA by a factor of 4.7$\pm$0.7~\cite{dAuCharm}) is
1.40$\pm$0.11$\pm$0.39 mb for central and 1.29$\pm$0.12$\pm$0.36
mb for minbias \AuAu\ collisions. The central values of the cross
sections reported by PHENIX~\cite{phenixcharmxsec} are a factor of
about two smaller than ours at all measured \pt. The difference is
approximately 1.5
times the combined uncertainties. %The FONLL
%calculation~\cite{vogt} shown as the band in
%Fig.~\ref{fig:figure4}, underpredicts the minbias data by a factor
%of $4.3\pm0.5{(stat.)}\pm1.2(syst.)_{-3.3}^{+9.4}(theory)$. The
%upper limit of the theory uncertainties is consistent with our
%result.
The FONLL calculation~\cite{vogt} is the band in
Fig.~\ref{fig:figure4}. The ratio of the minbias data over theory
calculation is
4.3$\pm$0.5$(stat.)$$\pm$1.2$(syst.)_{-3.3}^{+9.4}(theory)$. The
upper theory value reproduces our result.

%We note that a reevaluation of the theory uncertainties~\cite{vogt} may
%reduce the discrepancy between our result and pQCD predictions.

%Future detector upgrades will provide direct
%reconstruction of charmed hadrons and help quantify the radial
%flow and elliptic flow of charmed hadrons~\cite{HFT}.

%--====================================================================
%--====================================================================
%%%%%%%%%%%%%%%%%%%%%%%%%%%%%%%%%%%%%%%%%%%%%%%%%%%%%%%%%%%%%%%%%%%%%%
% SUMMARY
%%%%%%%%%%%%%%%%%%%%%%%%%%%%%%%%%%%%%%%%%%%%%%%%%%%%%%%%%%%%%%%%%%%%%%

In summary, we report measurements of charmed hadron production at
mid-rapidity from analysis of $D\rightarrow K\pi$ , $\mu$($e$)
from semileptonic charm decays in minbias and central \AuAu\
collisions at RHIC. The blast-wave fits and the direct comparisons
of the spectra suggest that charmed hadrons interact with and
decouple from the system differently from the light hadrons.  The
$d\sigma_{c\bar{c}}^{NN}/dy$ at mid-rapidity are extracted from a
combination of the three measurements covering the transverse
momentum range which contributes to $\sim$90\% of the integrated
cross section. The total cross sections are found to scale with
the number of binary collisions. This confirms the expected
scaling of hard production processes with binary interactions
among incoming nucleons so that charm quarks can be used as a
calibrated probe of the early-stage dynamics of the system.
%%%%%%%%%%%%%%%%%%%%%%%%%%%%%%%%%%%%%%%%%%%%%%%%%%%%%%%%%%%%%%%%%%%%%%
% ACKNOWLEDGMENT
%%%%%%%%%%%%%%%%%%%%%%%%%%%%%%%%%%%%%%%%%%%%%%%%%%%%%%%%%%%%%%%%%%%%%%

We thank the RHIC Operations Group and RCF at BNL, and the
NERSC Center at LBNL for their support. This work was supported
in part by the Offices of NP and HEP within the U.S. DOE Office
of Science; the U.S. NSF; the BMBF of Germany; CNRS/IN2P3, RA, RPL, and
EMN of France; EPSRC of the United Kingdom; FAPESP of Brazil;
the Russian Ministry of Sci. and Tech.; the Ministry of
Education and the NNSFC of China; IRP and GA of the Czech Republic,
FOM of the Netherlands, DAE, DST, and CSIR of the Government
of India; Swiss NSF; the Polish State Committee for Scientific
Research; Slovak Research and Development Agency, and the
Korea Sci. \& Eng. Foundation.


\begin{thebibliography}{}
\bibitem{muana} A stringent \dedx\ cut ($|n\sigma_{\pi}|<0.5$~\cite{ffcharm})
provides a pure pion sample. We then applied the muon $m^2$ window
and properly normalize this background $m^2$ distribution to the
inclusive sample at the pion mass peak. This provides the residual
pion contamination which was subtracted from the $\mu$ DCA
distribution.
\bibitem{fitBR}The averaged $c\rightarrow l$ B.R.=9.6\% was used
to normalize the spectra for the fit. The differences from
$c\rightarrow e$ (10.3\%) and $c\rightarrow \mu$ (8.2\%) were
taken into account in the final systematic errors.
\end{thebibliography}
\end{document}